\documentclass[%
prl,
amsmath,amssymb,
reprint,%
]{revtex4-2}

\usepackage{graphicx}
\usepackage{dcolumn}
\usepackage{bm}

\usepackage{textcomp}
\usepackage{siunitx}

\usepackage[T1]{fontenc}

\usepackage{booktabs}
\usepackage{amsmath}
\usepackage{graphicx}
\usepackage{dcolumn}
\usepackage{bm}
\usepackage{hyperref}
\usepackage[capitalize]{cleveref}
\newcommand{\rhoN}{\ensuremath{\rho_{\rm N}}}
\newcommand{\kB}{\ensuremath{k_{\rm B}}}

\newcommand{\citeref}[1]{\citeauthor{#1}\cite{#1}}

\newcommand{\LL}{\ensuremath{ \mathcal{L} }}

\usepackage{xcolor}
\usepackage{soul}

\newcommand{\papertitle}{Quantum Entropic Effects in the Liquid Viscosities of Hydrogen, Deuterium, and Neon}

\newcommand{\abstracttext}{The extremely low temperatures have limited the availability and accuracy of experimental thermophysical property measurements for cryogens, particularly transport properties. Traditional scaling techniques such as corresponding states theory have long been known to be inaccurate for fluids with strong quantum effects. To address this need, this paper investigates how quantum effects impact thermodynamics and momentum transfer (shear viscosity) in the fluid phases of hydrogen, deuterium, and neon. We utilize experimental viscosity measurements and reference empirical equations of state to show that conventional entropy scaling is inadequate for quantum-dominated systems. We then provide a simple empirical correction to entropy scaling based on the ratio of quantum to packing length scale that accounts for the deviations.}
\usepackage{cancel}
\usepackage{xcolor}
\usepackage{float}

\usepackage{zref-xr,zref-user}
\zxrsetup{verbose}
\zexternaldocument*{SI_quantumentropypaper}	

\graphicspath{{figs/}}

\begin{document}
	

%
\title{\papertitle}
%
\author{Ian H. Bell}
\email{ian.bell@nist.gov}
\thanks{Corresponding author}
\affiliation{%
Applied Chemicals and Materials Division, National Institute of Standards and Technology, Boulder, CO 80305, USA
}

\author{Jacob W. Leachman}
\affiliation{%
Hydrogen Properties for Energy Research Laboratory, School of Mechanical and Materials Engineering, Washington State University, Pullman, WA 99163, USA
}


\author{Albert F. Rigosi}
\affiliation{%
Quantum Measurement Division, National Institute of Standards and Technology, Gaithersburg, MD, 20899, USA 
}

\author{Heather M. Hill}
\affiliation{%
Quantum Measurement Division, National Institute of Standards and Technology, Gaithersburg, MD, 20899, USA
}

\date{\today}

\begin{abstract}
\abstracttext
\end{abstract}

\maketitle

\newcommand{\Supp}{SI}

The International Energy Agency (IEA) estimates that demand for green hydrogen will increase from 94 megatonnes annually in 2021 to 179.9 megatonnes in 2030 \cite{IEA-2022-H2}. 
Engineering systems to meet this need requires accurate knowledge of thermophysical property information, including transport properties like the shear viscosity. Few reference-quality measurements of cryogenic hydrogen shear viscosities are available, particularly for liquid-like states \cite{Leachman-IJT-2007}. Moreover, other cryogens ($^4$He, $^3$He, HD, D$_2$, HT, DT, T$_2$, and Ne) generally lack sufficient transport property information needed for reference quality transport property correlations. The very low shear viscosity of deep cryogens combined with the quantum nature of these fluids limits future prospects for robustly addressing this need via traditional empirical techniques. A better theoretical understanding of the transport properties of such systems is required. In order to avoid the extreme quantum effects present in $^4$He, $^3$He, this study limits its scope to H$_2$ and heavier  species.

In a classical two-parameter corresponding-states approach, thermodynamic properties can be expressed in terms of temperature and specific volume reduced by their critical-point values. Once quantum effects become relevant, a third parameter is needed, the thermal de Broglie wavelength defined by $\lambda_{\rm th} = h/\sqrt{2\pi m\kB T}$ where $h$ is Planck's constant, $m$ is the mass of one particle, $\kB$ is Boltzmann's constant, and $T$ is the temperature. Hydrogen has the lowest mass and therefore the largest de Broglie wavelength at any given temperature, so is in a sense the ``most quantum'' molecular species. \citeref{deBoer-PHYSICA-1938} highlights that the parameter of interest in dense phases is the thermal length scale relative to the packing length scale: $\LL\equiv\lambda_{\rm th}/\rhoN^{-1/3}$, where $\rhoN$ is the number density. This length-scale ratio quantifies the quantum effects; if $\LL \ll 1$, the physics are classically dominated. By this metric, liquid $^4$He is ``more quantum'' than liquid hydrogen because while their liquid densities are similar, the de Broglie wavelength of liquid $^4$He is larger because its temperature is significantly lower (see \cref{tab:small_info}). In dilute gases, the relevant length scale is the de Broglie wavelength relative to the molecular size, which was considered by Ref. \citenum{deBoer-PHYSICA-1938}.

Isomorph theory \cite{Bailey-JCP-2008-PartI, Bailey-JCP-2008-PartII, Gnan-JCP-2009-PartIV, Schroder-JCP-2009-PartIII, Schroder-JCP-2011-PartV} explains how the macroscopically scaled transport properties should be a monovariate function of the excess entropy over much of the phase diagram for many classical fluids \cite{Dyre-JCP-2018-Review}, especially in the liquid phase. The macroscopically scaled viscosity is defined by  $\widetilde{\eta} = \eta/(\rhoN^{2/3}\sqrt{m\kB T})$ where $\eta$ is the shear viscosity. The excess entropy per particle is defined as the entropy of a fluid minus that of the non-interacting ideal gas at the same temperature and density: $s_{\rm ex} \equiv s(T,\rho) - s^{\rm(ig)}(T,\rho)$.  For convenience and conceptual understanding, the non-dimensional quantity $s^+$ is defined by $s^+\equiv-s_{\rm ex}/\kB$, a quantity which must always be positive in general, and increases as the fluid becomes more ``structured'' (microstates of the system are made inaccessible relative to the non-interacting reference system). The values of $s^+$ are obtained from equations of state (EOS), as detailed in \S\zref{sec:EOS} in the SI. The excess entropy coming from an EOS captures all the quantum and classical physics - experimental measurements used to develop the EOS do not allow for the two to be disentangled. The value of $\LL$ quantifies the relative influence of quantum effects at a state point and motivates the investigation of how quantum effects alter $s^+$ from its classical value. The ideal-gas contribution for a monatomic classical species is $s^{\rm (ig,cl,mono)}=5/2-\ln(\mathcal{L}^3)$, and a power series in terms of $\mathcal{L}$ represents the corrections to the classical formulation (see \S\zref{sec:idealcorrection} in the SI)

The existence of a monovariate relationship of unknown form between $\widetilde{\eta}$ and $s^+$ is an exact solution for classical inverse-power-law pair potential fluids over the entire fluid phase diagram (the hard sphere being the limiting case of this family)\cite{Dyre-JPCB-2014-hidden}. The transport properties of the Lennard-Jones fluid (the most studied model system for a fluid with both attraction and repulsion) can be represented within their simulation uncertainty by a simple entropy scaling approach \cite{Bell-JPCB-2019-LJ} and other simple \cite{Polychroniadou-IJT-2021-Kr} and not so simple fluids \cite{Yang-JCED-2021-etarefrig,Young-JCP-2023-salts} can be modeled by a similar approach.  The review of Ref. \citenum{Dyre-JCP-2018-Review} covers many of the applications of this approach up to the year 2018. 

\begin{figure}
	\includegraphics[width=3.3in]{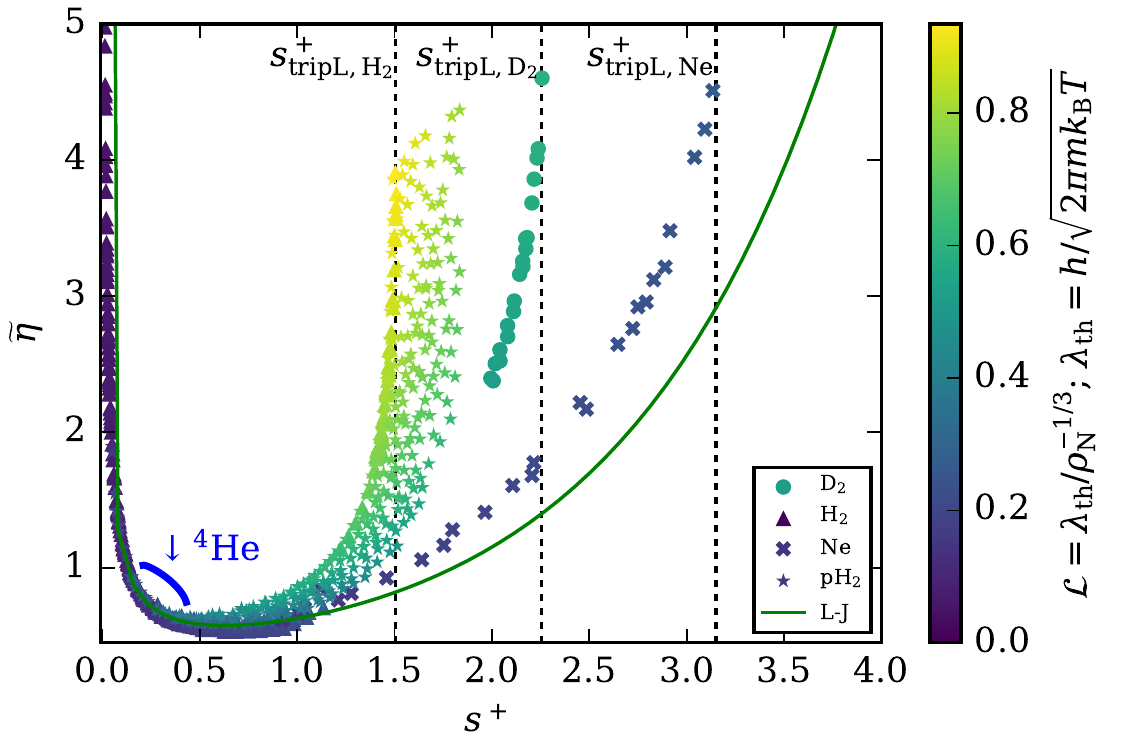}
	\caption{Conventional entropy scaling applied to the experimental shear viscosity data of cryogens.  Each marker corresponds to an experimental data point and the values of $\rhoN$ and $s^+$ are obtained from the respective reference EOS. Values calculated for saturated liquid helium from the empirical models in REFPROP 10.0 \cite{Arp-NIST-1998} are also plotted. \label{fig:tildeeta_H2}}
\end{figure}

The experimental data for the cryogens were largely captured in the 1960s and 1970s during the Space Race. The study of Refs. \citenum{Muzny-JCED-2013,Muzny-JCED-2022} includes the literature sources for normal hydrogen (H$_2$) and parahydrogen (pH$_2$) available at the time, and they built an empirical correlation for the viscosity that reproduced the liquid-phase viscosity generally within 4\%. The collected experimental data for hydrogen and other cryogens are plotted in \cref{fig:tildeeta_H2} in macroscopically scaled form, colored according to the length-scale ratio. Further discussion of the selected datasets is available in \S\zref{sec:expdata} in the SI. The primary datasets selected by Ref. \citenum{Muzny-JCED-2013} were used for hydrogen, and the datasets used in developing the model in NIST REFPROP \cite{Huber-NISTIR-2018} were used for neon.  For neon and deuterium, the only reasonable experimental data in the liquid phase at cryogenic temperatures are for saturated liquid states. To highlight the wildly different nature of helium, the data for saturated liquid $^4$He were plotted with the default models in NIST REFPROP \cite{LEMMON-RP10}. The conventional quantum corresponding-states approach of Ref. \citenum{Rogers-PHYSICA-1966} yields a similar result. While the data for $s^+ < 0.5$ mostly follow a single master curve, the liquid phase ($s^+ > 0.6$) data for H$_2$ and pH$_2$ do not. The deviations from monovariability are linked to $\LL$, an observation exploited in this work to develop a modified version of entropy scaling accounting for quantum effects. The curve plotted for Lennard-Jones is $\widetilde{\eta} = 0.2163 \exp(1.068s^+)/(s^+)^{2/3}$ and comes from Ref. \citenum{Bell-JPCB-2019-LJ}; this curve can be taken as a reasonable classical limit.

\begin{table}
	\caption{Selected values for the cryogens studied here according to the reference EOS indicated in the reference. \label{tab:small_info}}
	\begin{tabular}{lrrrl}
\toprule
                          name & $M$ / g/mol & $T_{\rm trip}$ / K & $s^+_{\rm tripL}$ & $\mathcal{L}_{\rm tripL}=\lambda_{\rm th}/\rho_{\rm N}^{-1/3}$ \\
\midrule
                        $^4$He &       4.003 &               2.18 &             0.208 &                                  1.656=0.591/0.357 \\
pH$_2$\cite{10.1063/1.3160306} &       2.016 &              13.80 &             1.493 &                                  0.941=0.331/0.352 \\
 H$_2$\cite{10.1063/1.3160306} &       2.016 &              13.96 &             1.504 &                                  0.936=0.329/0.352 \\
 D$_2$\cite{10.1063/1.4864752} &       4.028 &              18.72 &             2.256 &                                  0.596=0.201/0.337 \\
                            Ne &      20.179 &              24.56 &             3.151 &                                  0.262=0.078/0.299 \\
     Ar\cite{10.1063/1.556037} &      39.948 &              83.81 &             3.553 &                                  0.084=0.030/0.360 \\
    Kr\cite{10.1021/je050186n} &      83.798 &             115.78 &             3.547 &                                  0.046=0.018/0.385 \\
    Xe\cite{10.1021/je050186n} &     131.293 &             161.41 &             3.575 &                                  0.029=0.012/0.419 \\
\bottomrule
\end{tabular}
\\
	$\dagger$: At the lambda temperature
\end{table}

Ref. \citenum{Andrade-NATURE-1931} proposed a universal value of $\widetilde{\eta}=5.47$ at the normal melting point; while the heavy noble gases do not have the same value in the liquid phase at the triple point (hereafter we drop the liquid phase modifier but it remains implied if not stated), their value is quite consistent at $4.35\pm0.4$ (see \S\zref{sec:neonanalysis} in the SI), even hydrogen is in the same range. The expression for $\widetilde{\eta}$ can be re-written in an evocative form:
\begin{equation}
	\widetilde{\eta}=\frac{\eta}{\rhoN}\mathcal{L}\frac{\sqrt{2\pi}}{h}
\end{equation}
in terms of a kinematic-viscosity-like term ($\eta/\rhoN$), the length-scale ratio $\mathcal{L}$, and a proportionality constant. From the values at the respective triple point, $\eta/\rhoN$ is very nearly inversely proportional to $\mathcal{L}$ and thus their product is nearly constant.

\begin{figure}
	\includegraphics[width=3.5in]{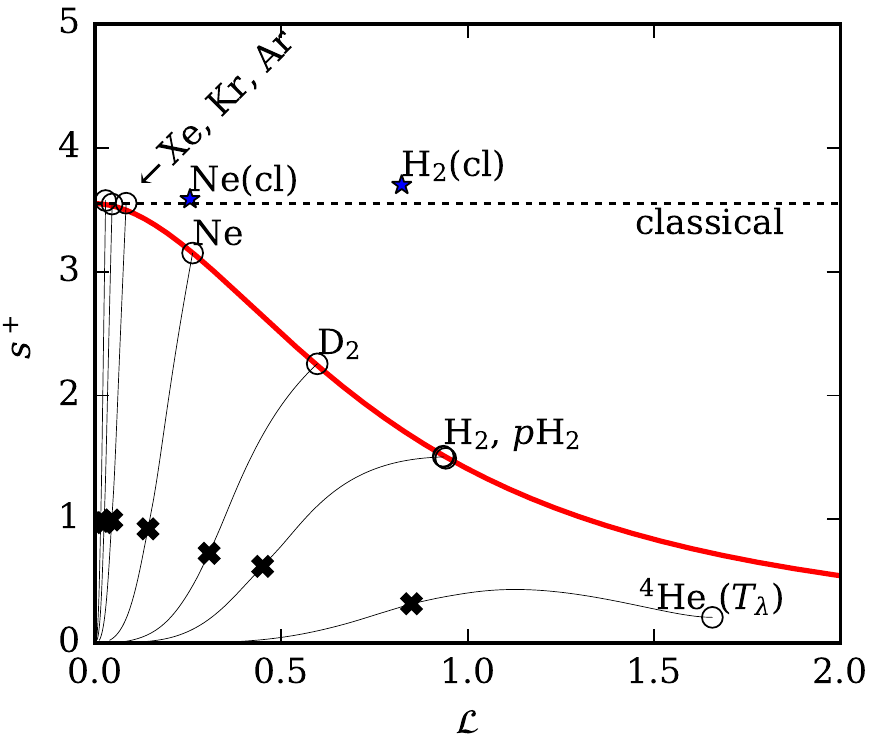}
	\caption{Values of $s^+$ for orthobaric densities (thin curves), in the liquid phase at the liquid-vapor-solid triple point (open circle) and the critical point (filled X) for cryogens as a function of $\LL$ according to the respective reference EOS.  For helium, the ``triple point'' value pertains to the liquid phase at the lambda temperature of 2.1768 K.  The thick red curve is the quantum correction from \cref{eq:splusrescaler}. The stars correspond to the values obtained from classical calculations including two- and three-body interactions. \label{fig:offset_values}}
\end{figure}

The results at the triple point suggest that the scatter in \cref{fig:tildeeta_H2} does not appear to be caused by deviations in the vertical ($\widetilde{\eta}$) direction caused by quantum effects. Rather, the dominant effect is that quantum effects shift the data in the $s^+$ direction. In order to investigate how the values of $s^+$ are influenced by quantum effects, the values of $s^+$ of the liquid phase at the solid-liquid-vapor triple point are considered. In a classical corresponding-states approach we would expect the value to be the same for all species that have conformal pairwise potentials. \Cref{fig:offset_values} shows the $s^+$ values for the cryogens, as well as values calculated along the co-existence curve and at the vapor-liquid critical point. The heavy noble gases (Ar, Kr, Xe) are clustered together at 3.55 and the lighter species have values below those of the heavy noble gases.  

The conventional understanding of quantum effects is based upon ``quantum swelling'' \cite[page 322]{Chapman-BOOK-1970} caused by the spatial delocalization of the centers of mass of the atoms, as appears in a number of thermodynamic modeling approaches accounting for quantum effects \cite{Aasen-JCP-2019,Aasen-JCP-2020,Aasen-FPE-2020, Walker-JCP-2022,Yoon-JCP-1988,Deiters-FPE-1983}. In this approach, a larger co-volume of the atom or molecule caused by quantum effects results in a larger value of $s^+$ (the phase is more structured). As an illustration, the venerable yet simple van der Waals EOS yields $s^+_{\rm vdW} = -\ln(1-b\rhoN)$, thus an increase in the co-volume $b$ due to quantum swelling increases $s^+$ at constant density. Similar empirical approaches are outlined in  \S\zref{sec:realcorrection} in the SI.  On the contrary (see \cref{fig:offset_values}), triple-point values of $s^+$ for quantum fluids are smaller than their classical limit so quantum swelling alone cannot account for the shift. 

A rigorous mechanism for the reduction in $s^+$ in condensed phases is out of reach at present (a \textit{post hoc} rationalization is below). With a few assumptions, the impact of quantum effects can be straightforwardly quantified, allowing for a powerful predictive model and hints that a new theoretical understanding is possible. The first assumption is that the triple point value of $s^+$ \textit{should} be 3.55 for classical neon, deuterium, and hydrogen, in agreement with the heavy noble gases (Ar, Kr, Xe).  Then in a second step we assume that the same relationship between quantum and classical values of $s^+$ and $\LL$ that holds at the triple point also holds everywhere in the phase diagram.  

\begin{figure}
	\includegraphics[width=3.2in]{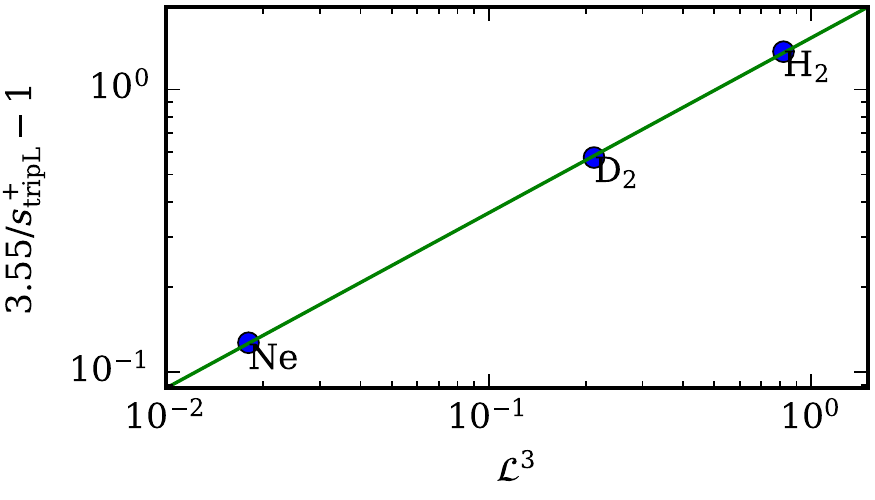}
	\caption{Ratio of hypothesized classical value of $s^+$ at liquid phase of SLV triple point to quantum value as a function of $\mathcal{L}^3$ for cryogens with meaningful quantum effects and a solid-vapor-liquid triple point\label{fig:expcorrection_plot}}
\end{figure}

Values of the ratio of 3.55 to the value of $s^+$ at the triple point are shown in \cref{fig:expcorrection_plot} as a function of $\LL^3$. While the fitted curve is based upon empirical EOS for only three fluids, the nearly perfect power-law relationship is highly suggestive. With this curve, the first assumption allows for an empirical relationship linking the ratio to $\mathcal{L}^3$ of the form
\begin{equation}
	\label{eq:splusrescaler}
	s^+_{\rm cl} = s^+_{\rm qu}(1+a(\mathcal{L}^3)^b)
\end{equation}
with $a=1.5232$ and $b=0.6198$. The values of $s^+_{\rm qu}$ are reproduced within $\pm 0.12$, which is likely within the uncertainty of the EOS, and the curve from \cref{eq:splusrescaler} is plotted in \cref{fig:offset_values}, with the value of $s^+_{\rm cl}$ taken to be 3.55.

In order to test the hypothesis that the liquid triple-point value of classical monatomic species should be $s^+=3.55$, calculations for classical Ne and H$_2$ (without quantum corrections) with two-body and three-body interactions included were carried out following the approach in Refs. \citenum{Deiters-JPCB-2020,Deiters-JPCB-2021}. The method yields values for Ne and H$_2$ of 3.59 and 3.70 at the respective classical triple point (which is not the same as the quantum one). Considering the accuracy of the pair potential, the accuracy of the three-body interactions, the neglect of more than three-body interactions, and other sources of uncertainty, these simulation results appear to confirm the hypothesis. In a classical calculation, there is no distinction between pH$_2$, H$_2$, and D$_2$. The method is described in \S\zref{sec:classical} of the SI, including a detailed description of pre-publication results from Ref. \citenum{Sadus-JCP-2023}.

\begin{figure}
	\includegraphics[width=3.3in]{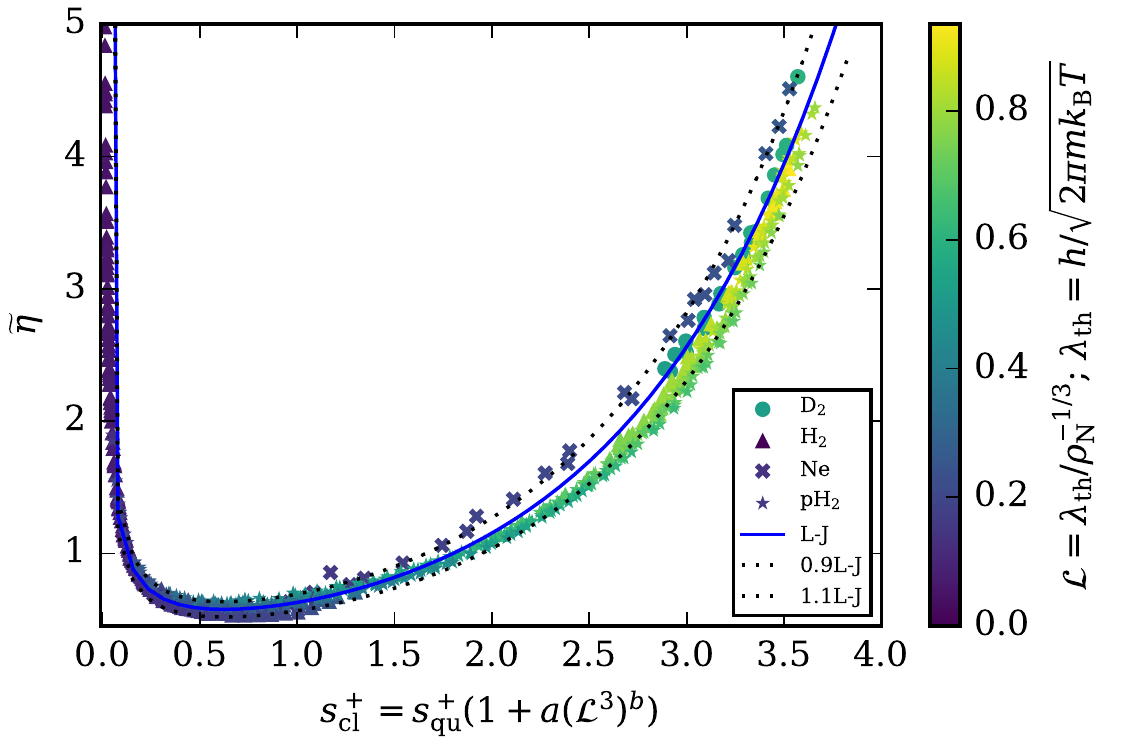}
	\caption{Macroscopic scaling applied to the shear viscosity of cryogens as a function of quantum-corrected values of $s^+$.  The dashed curves indicate $\pm$10\% relative to the correlation for Lennard-Jones. \label{fig:expcorrection_entropy_scaling}}
\end{figure}

Further evidence of the success of this empirical approach is presented in \cref{fig:expcorrection_entropy_scaling}. The experimental viscosity data are plotted with their values of $s^+$ re-scaled according to \cref{eq:splusrescaler} to yield a quantum corrected pseudo-classical value; the curve for Lennard-Jones is not rescaled. After the correction for quantum effects, the data fall very close to the curve for the Lennard-Jones fluid \cite{Bell-JPCB-2019-LJ}. There is a systematic bias for neon agreeing with the offset in $\widetilde{\eta}$ at the triple point (other experimental viscosity data for neon show significant scatter\cite{Herreman-CRYO-1974}) as shown in \S\zref{sec:neonanalysis} in the SI. Overall the collapse of the data after applying the entropy correction is surprisingly good, indicating that the simple correction is capturing much of the right physics.

\begin{figure}
	\includegraphics[width=3.3in]{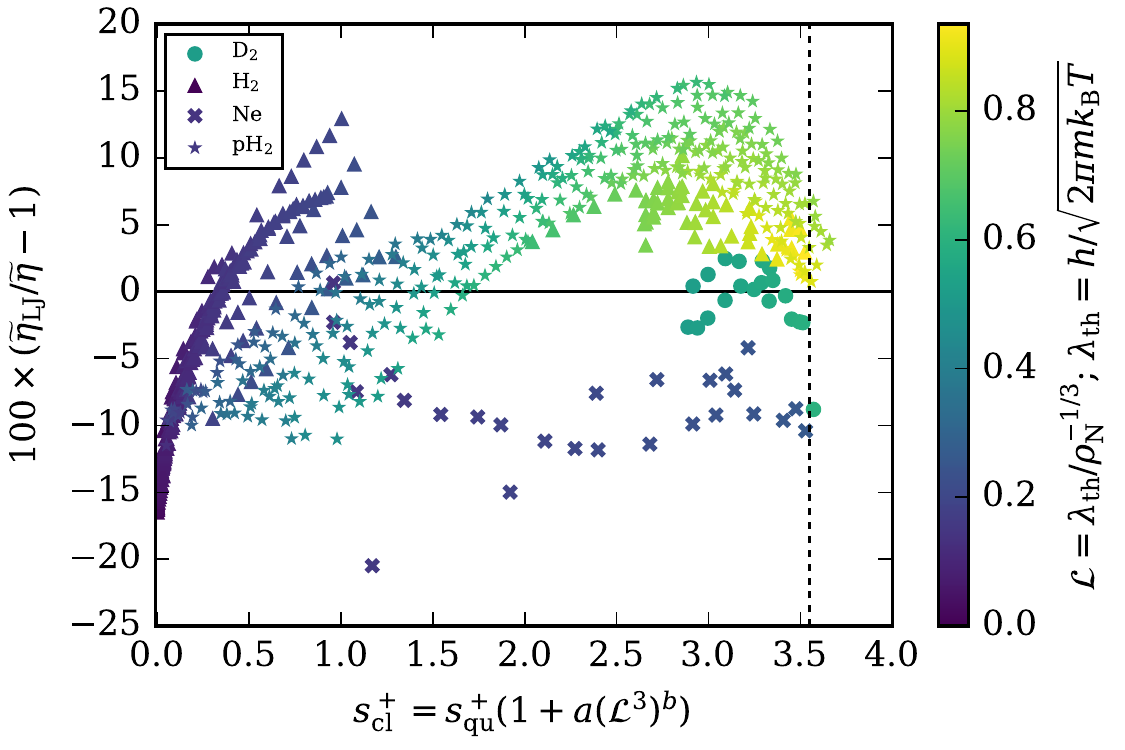}
	\caption{Deviations of shear viscosity points from the correlation for Lennard-Jones in Ref. \citenum{Bell-JPCB-2019-LJ}.  The vertical dashed curve indicates the classical value of 3.55. \label{fig:expcorrection_entropy_scaling_deviations}}
\end{figure}

To quantify the success of this method,  deviations between the scaled data points and the correlation for the Lennard-Jones fluid are shown in \cref{fig:expcorrection_entropy_scaling_deviations}. Even if the Lennard-Jones model is not a perfect analogy for noble gases \cite{Rutkai-MP-2016}, it captures many of the correct qualitative features.  Most rescaled points agree with the Lennard-Jones empirical correlation to within $\pm 10$\%. Neon demonstrates a systematic offset of -10\%, and the limited deuterium data agree well with the Lennard-Jones curve. Other quantum fluids (especially technically relevant isomers of hydrogen) are expected to fit within the same scaling framework. In light of the significant scatter in the experimental liquid viscosity data of liquid cryogens, this simple approach is remarkably successful.

In order to understand the physics of the correction, to a good approximation, the value of $s^+$ at the triple point (see \S\zref{sec:splusmath} in the SI for the derivation) can be given by
\begin{equation}
s^+_{\rm trip L} \approx 1-\frac{u^{\rm ex}_{\rm L}}{\kB T}-\ln\left(\frac{\rho_{\rm L}}{\rho_{\rm V}}\right)
\end{equation}
highlighting two contributions, one from the potential energy of the ensemble of molecules ($u^{\rm ex}_{\rm L}$ is a measure of all \textit{inter}molecular interactions), and another from the density ratio (see \S\zref{sec:splusmath} in the SI). The first term $u^{\rm ex}_{\rm L}/(\kB T)$ has a functional dependence on $\LL^3$ similar to $s^+$, though not following a perfect power-law, while $ \ln\left(\rho_{\rm L}/\rho_{\rm V}\right)$ has a more uneven behavior. It is only when the terms are combined in $s^+$ that the power law relationship occurs, highlighting the power of thinking in terms of $s^+$ rather than other related thermodynamic quantities.

Ref. \citenum{Nagashima-JCP-2017} observed that the quantum effects have two competing impacts: they simultaneously increase the particle diameter and decrease the effective depth of the attractive well of the potential. The relative impact of these two effects is state-point dependent, and our results suggest that in the dense liquid phase of cryogens, the reduction in the interaction well depth is the dominant effect. This has been quantified also in Ref. \citenum{Walker-JCP-2022} where they note a very similar reduction in the depth of the attractive well as the relative quantumness increases (see \S\zref{sec:Walker} in the SI). At the triple point, hydrogen has a reduction in $s^+$ of approximately 2 from the classical value, of which $1.5$ comes from the internal energy, and $0.5$ from the density ratio.

The origins of this work were an attempt to apply the insights of isomorph theory to the transport properties of quantum-influenced fluids. Progress in that problem largely rests on the question of how quantum effects impact the entropy of the fluid, and correcting back to a pseudo-classical excess entropy recovers the classical behavior of the transport properties to a good approximation. The best equations of state in existence today suggest that a novel and simple understanding of how quantum effects influence the thermodynamic properties of molecular systems is within reach. This new knowledge offers a means of probing the fundamental nature of interactions and structure in liquid systems. We invite readers intrigued by these questions to search for a more rigorous explanation.

    In order to ensure reproducibility of our results, the supplementary material includes: extended derivations and discussions, additional figures, tables of experimental data sources, and the classical simulation information.

    \begin{acknowledgments}
   	The authors gratefully acknowledge Erin Espeland's contributions to early calculations used in this work. Thanks go to Ulrich Deiters and Richard Sadus for making their classical calculations available prior to their publication. Discussions with Vincent Arp, Allan Harvey, Pierre Walker, Karl Irikura, {\O}ivind Wilhelmsen, Giovanni Garberoglio, Dan Friend, Tianpu Zhao, and Luis Pedro Garcia-Pintos, provided helpful insights.
    \end{acknowledgments}


\begin{thebibliography}{38}%
\makeatletter
\providecommand \@ifxundefined [1]{%
 \@ifx{#1\undefined}
}%
\providecommand \@ifnum [1]{%
 \ifnum #1\expandafter \@firstoftwo
 \else \expandafter \@secondoftwo
 \fi
}%
\providecommand \@ifx [1]{%
 \ifx #1\expandafter \@firstoftwo
 \else \expandafter \@secondoftwo
 \fi
}%
\providecommand \natexlab [1]{#1}%
\providecommand \enquote  [1]{``#1''}%
\providecommand \bibnamefont  [1]{#1}%
\providecommand \bibfnamefont [1]{#1}%
\providecommand \citenamefont [1]{#1}%
\providecommand \href@noop [0]{\@secondoftwo}%
\providecommand \href [0]{\begingroup \@sanitize@url \@href}%
\providecommand \@href[1]{\@@startlink{#1}\@@href}%
\providecommand \@@href[1]{\endgroup#1\@@endlink}%
\providecommand \@sanitize@url [0]{\catcode `\\12\catcode `\$12\catcode
  `\&12\catcode `\#12\catcode `\^12\catcode `\_12\catcode `\%12\relax}%
\providecommand \@@startlink[1]{}%
\providecommand \@@endlink[0]{}%
\providecommand \url  [0]{\begingroup\@sanitize@url \@url }%
\providecommand \@url [1]{\endgroup\@href {#1}{\urlprefix }}%
\providecommand \urlprefix  [0]{URL }%
\providecommand \Eprint [0]{\href }%
\providecommand \doibase [0]{https://doi.org/}%
\providecommand \selectlanguage [0]{\@gobble}%
\providecommand \bibinfo  [0]{\@secondoftwo}%
\providecommand \bibfield  [0]{\@secondoftwo}%
\providecommand \translation [1]{[#1]}%
\providecommand \BibitemOpen [0]{}%
\providecommand \bibitemStop [0]{}%
\providecommand \bibitemNoStop [0]{.\EOS\space}%
\providecommand \EOS [0]{\spacefactor3000\relax}%
\providecommand \BibitemShut  [1]{\csname bibitem#1\endcsname}%
\let\auto@bib@innerbib\@empty
\bibitem [{\citenamefont {IEA}(2022)}]{IEA-2022-H2}%
  \BibitemOpen
  \bibfield  {author} {\bibinfo {author} {\bibnamefont {IEA}},\ }\href
  {https://www.iea.org/reports/hydrogen} {\emph {\bibinfo {title}
  {Hydrogen}}},\ \bibinfo {type} {Tech. Rep.}\ (\bibinfo  {institution} {IEA},\
  \bibinfo {year} {2022})\BibitemShut {NoStop}%
\bibitem [{\citenamefont {Leachman}\ \emph {et~al.}(2007)\citenamefont
  {Leachman}, \citenamefont {Jacobsen}, \citenamefont {Penoncello},\ and\
  \citenamefont {Huber}}]{Leachman-IJT-2007}%
  \BibitemOpen
  \bibfield  {author} {\bibinfo {author} {\bibfnamefont {J.~W.}\ \bibnamefont
  {Leachman}}, \bibinfo {author} {\bibfnamefont {R.~T.}\ \bibnamefont
  {Jacobsen}}, \bibinfo {author} {\bibfnamefont {S.~G.}\ \bibnamefont
  {Penoncello}},\ and\ \bibinfo {author} {\bibfnamefont {M.~L.}\ \bibnamefont
  {Huber}},\ }\bibfield  {title} {\bibinfo {title} {{Current Status of
  Transport Properties of Hydrogen}},\ }\href
  {https://doi.org/10.1007/s10765-007-0229-4} {\bibfield  {journal} {\bibinfo
  {journal} {Int. J. Thermophys.}\ }\textbf {\bibinfo {volume} {28}},\ \bibinfo
  {pages} {773} (\bibinfo {year} {2007})}\BibitemShut {NoStop}%
\bibitem [{\citenamefont {de~Boer}\ and\ \citenamefont
  {Michels}(1938)}]{deBoer-PHYSICA-1938}%
  \BibitemOpen
  \bibfield  {author} {\bibinfo {author} {\bibfnamefont {J.}~\bibnamefont
  {de~Boer}}\ and\ \bibinfo {author} {\bibfnamefont {A.}~\bibnamefont
  {Michels}},\ }\bibfield  {title} {\bibinfo {title} {Contribution to the
  quantum-mechanical theory of the equation of state and the law of
  corresponding states. determination of the law of force of helium},\ }\href
  {https://doi.org/10.1016/s0031-8914(38)80037-9} {\bibfield  {journal}
  {\bibinfo  {journal} {Physica}\ }\textbf {\bibinfo {volume} {5}},\ \bibinfo
  {pages} {945} (\bibinfo {year} {1938})}\BibitemShut {NoStop}%
\bibitem [{\citenamefont {Bailey}\ \emph
  {et~al.}(2008{\natexlab{a}})\citenamefont {Bailey}, \citenamefont {Pedersen},
  \citenamefont {Gnan}, \citenamefont {Schr{\o}der},\ and\ \citenamefont
  {Dyre}}]{Bailey-JCP-2008-PartI}%
  \BibitemOpen
  \bibfield  {author} {\bibinfo {author} {\bibfnamefont {N.~P.}\ \bibnamefont
  {Bailey}}, \bibinfo {author} {\bibfnamefont {U.~R.}\ \bibnamefont
  {Pedersen}}, \bibinfo {author} {\bibfnamefont {N.}~\bibnamefont {Gnan}},
  \bibinfo {author} {\bibfnamefont {T.~B.}\ \bibnamefont {Schr{\o}der}},\ and\
  \bibinfo {author} {\bibfnamefont {J.~C.}\ \bibnamefont {Dyre}},\ }\bibfield
  {title} {\bibinfo {title} {{Pressure-energy correlations in liquids. I.
  Results from computer simulations}},\ }\href
  {https://doi.org/10.1063/1.2982247} {\bibfield  {journal} {\bibinfo
  {journal} {J. Chem. Phys.}\ }\textbf {\bibinfo {volume} {129}},\ \bibinfo
  {pages} {184507} (\bibinfo {year} {2008}{\natexlab{a}})}\BibitemShut
  {NoStop}%
\bibitem [{\citenamefont {Bailey}\ \emph
  {et~al.}(2008{\natexlab{b}})\citenamefont {Bailey}, \citenamefont {Pedersen},
  \citenamefont {Gnan}, \citenamefont {Schr{\o}der},\ and\ \citenamefont
  {Dyre}}]{Bailey-JCP-2008-PartII}%
  \BibitemOpen
  \bibfield  {author} {\bibinfo {author} {\bibfnamefont {N.~P.}\ \bibnamefont
  {Bailey}}, \bibinfo {author} {\bibfnamefont {U.~R.}\ \bibnamefont
  {Pedersen}}, \bibinfo {author} {\bibfnamefont {N.}~\bibnamefont {Gnan}},
  \bibinfo {author} {\bibfnamefont {T.~B.}\ \bibnamefont {Schr{\o}der}},\ and\
  \bibinfo {author} {\bibfnamefont {J.~C.}\ \bibnamefont {Dyre}},\ }\bibfield
  {title} {\bibinfo {title} {{Pressure-energy correlations in liquids. II.
  Analysis and consequences}},\ }\href {https://doi.org/10.1063/1.2982249}
  {\bibfield  {journal} {\bibinfo  {journal} {J. Chem. Phys.}\ }\textbf
  {\bibinfo {volume} {129}},\ \bibinfo {pages} {184508} (\bibinfo {year}
  {2008}{\natexlab{b}})}\BibitemShut {NoStop}%
\bibitem [{\citenamefont {Gnan}\ \emph {et~al.}(2009)\citenamefont {Gnan},
  \citenamefont {Schr{\o}der}, \citenamefont {Pedersen}, \citenamefont
  {Bailey},\ and\ \citenamefont {Dyre}}]{Gnan-JCP-2009-PartIV}%
  \BibitemOpen
  \bibfield  {author} {\bibinfo {author} {\bibfnamefont {N.}~\bibnamefont
  {Gnan}}, \bibinfo {author} {\bibfnamefont {T.~B.}\ \bibnamefont
  {Schr{\o}der}}, \bibinfo {author} {\bibfnamefont {U.~R.}\ \bibnamefont
  {Pedersen}}, \bibinfo {author} {\bibfnamefont {N.~P.}\ \bibnamefont
  {Bailey}},\ and\ \bibinfo {author} {\bibfnamefont {J.~C.}\ \bibnamefont
  {Dyre}},\ }\bibfield  {title} {\bibinfo {title} {{Pressure-energy
  correlations in liquids. IV. ``Isomorphs" in liquid phase diagrams}},\ }\href
  {https://doi.org/10.1063/1.3265957} {\bibfield  {journal} {\bibinfo
  {journal} {J. Chem. Phys.}\ }\textbf {\bibinfo {volume} {131}},\ \bibinfo
  {pages} {234504} (\bibinfo {year} {2009})}\BibitemShut {NoStop}%
\bibitem [{\citenamefont {Schr{\o}der}\ \emph {et~al.}(2009)\citenamefont
  {Schr{\o}der}, \citenamefont {Bailey}, \citenamefont {Pedersen},
  \citenamefont {Gnan},\ and\ \citenamefont
  {Dyre}}]{Schroder-JCP-2009-PartIII}%
  \BibitemOpen
  \bibfield  {author} {\bibinfo {author} {\bibfnamefont {T.~B.}\ \bibnamefont
  {Schr{\o}der}}, \bibinfo {author} {\bibfnamefont {N.~P.}\ \bibnamefont
  {Bailey}}, \bibinfo {author} {\bibfnamefont {U.~R.}\ \bibnamefont
  {Pedersen}}, \bibinfo {author} {\bibfnamefont {N.}~\bibnamefont {Gnan}},\
  and\ \bibinfo {author} {\bibfnamefont {J.~C.}\ \bibnamefont {Dyre}},\
  }\bibfield  {title} {\bibinfo {title} {{Pressure-energy correlations in
  liquids. {III}. Statistical mechanics and thermodynamics of liquids with
  hidden scale invariance}},\ }\href {https://doi.org/10.1063/1.3265955}
  {\bibfield  {journal} {\bibinfo  {journal} {J. Chem. Phys.}\ }\textbf
  {\bibinfo {volume} {131}},\ \bibinfo {pages} {234503} (\bibinfo {year}
  {2009})}\BibitemShut {NoStop}%
\bibitem [{\citenamefont {Schr{\o}der}\ \emph {et~al.}(2011)\citenamefont
  {Schr{\o}der}, \citenamefont {Gnan}, \citenamefont {Pedersen}, \citenamefont
  {Bailey},\ and\ \citenamefont {Dyre}}]{Schroder-JCP-2011-PartV}%
  \BibitemOpen
  \bibfield  {author} {\bibinfo {author} {\bibfnamefont {T.~B.}\ \bibnamefont
  {Schr{\o}der}}, \bibinfo {author} {\bibfnamefont {N.}~\bibnamefont {Gnan}},
  \bibinfo {author} {\bibfnamefont {U.~R.}\ \bibnamefont {Pedersen}}, \bibinfo
  {author} {\bibfnamefont {N.~P.}\ \bibnamefont {Bailey}},\ and\ \bibinfo
  {author} {\bibfnamefont {J.~C.}\ \bibnamefont {Dyre}},\ }\bibfield  {title}
  {\bibinfo {title} {{Pressure-energy correlations in liquids. V. Isomorphs in
  generalized Lennard-Jones systems}},\ }\href
  {https://doi.org/10.1063/1.3582900} {\bibfield  {journal} {\bibinfo
  {journal} {J. Chem. Phys.}\ }\textbf {\bibinfo {volume} {134}},\ \bibinfo
  {pages} {164505} (\bibinfo {year} {2011})}\BibitemShut {NoStop}%
\bibitem [{\citenamefont {Dyre}(2018)}]{Dyre-JCP-2018-Review}%
  \BibitemOpen
  \bibfield  {author} {\bibinfo {author} {\bibfnamefont {J.~C.}\ \bibnamefont
  {Dyre}},\ }\bibfield  {title} {\bibinfo {title} {{Perspective: Excess-entropy
  scaling}},\ }\href {https://doi.org/10.1063/1.5055064} {\bibfield  {journal}
  {\bibinfo  {journal} {J. Chem. Phys.}\ }\textbf {\bibinfo {volume} {149}},\
  \bibinfo {pages} {210901} (\bibinfo {year} {2018})}\BibitemShut {NoStop}%
\bibitem [{\citenamefont {Dyre}(2014)}]{Dyre-JPCB-2014-hidden}%
  \BibitemOpen
  \bibfield  {author} {\bibinfo {author} {\bibfnamefont {J.~C.}\ \bibnamefont
  {Dyre}},\ }\bibfield  {title} {\bibinfo {title} {{Hidden Scale Invariance in
  Condensed Matter}},\ }\href {https://doi.org/10.1021/jp501852b} {\bibfield
  {journal} {\bibinfo  {journal} {J. Phys. Chem. B}\ }\textbf {\bibinfo
  {volume} {118}},\ \bibinfo {pages} {10007} (\bibinfo {year}
  {2014})}\BibitemShut {NoStop}%
\bibitem [{\citenamefont {Bell}\ \emph {et~al.}(2019)\citenamefont {Bell},
  \citenamefont {Messerly}, \citenamefont {Thol}, \citenamefont {Costigliola},\
  and\ \citenamefont {Dyre}}]{Bell-JPCB-2019-LJ}%
  \BibitemOpen
  \bibfield  {author} {\bibinfo {author} {\bibfnamefont {I.~H.}\ \bibnamefont
  {Bell}}, \bibinfo {author} {\bibfnamefont {R.}~\bibnamefont {Messerly}},
  \bibinfo {author} {\bibfnamefont {M.}~\bibnamefont {Thol}}, \bibinfo {author}
  {\bibfnamefont {L.}~\bibnamefont {Costigliola}},\ and\ \bibinfo {author}
  {\bibfnamefont {J.~C.}\ \bibnamefont {Dyre}},\ }\bibfield  {title} {\bibinfo
  {title} {{Modified Entropy Scaling of the Transport Properties of the
  Lennard-Jones Fluid}},\ }\href {https://doi.org/10.1021/acs.jpcb.9b05808}
  {\bibfield  {journal} {\bibinfo  {journal} {J. Phys. Chem. B}\ }\textbf
  {\bibinfo {volume} {123}},\ \bibinfo {pages} {6345} (\bibinfo {year}
  {2019})}\BibitemShut {NoStop}%
\bibitem [{\citenamefont {Polychroniadou}\ \emph {et~al.}(2021)\citenamefont
  {Polychroniadou}, \citenamefont {Antoniadis}, \citenamefont {Assael},\ and\
  \citenamefont {Bell}}]{Polychroniadou-IJT-2021-Kr}%
  \BibitemOpen
  \bibfield  {author} {\bibinfo {author} {\bibfnamefont {S.}~\bibnamefont
  {Polychroniadou}}, \bibinfo {author} {\bibfnamefont {K.~D.}\ \bibnamefont
  {Antoniadis}}, \bibinfo {author} {\bibfnamefont {M.~J.}\ \bibnamefont
  {Assael}},\ and\ \bibinfo {author} {\bibfnamefont {I.~H.}\ \bibnamefont
  {Bell}},\ }\bibfield  {title} {\bibinfo {title} {{A Reference Correlation for
  the Viscosity of Krypton From Entropy Scaling}},\ }\href
  {https://doi.org/10.1007/s10765-021-02927-5} {\bibfield  {journal} {\bibinfo
  {journal} {Int. J. Thermophys.}\ }\textbf {\bibinfo {volume} {43}},\ \bibinfo
  {pages} {6} (\bibinfo {year} {2021})}\BibitemShut {NoStop}%
\bibitem [{\citenamefont {Yang}\ \emph {et~al.}(2021)\citenamefont {Yang},
  \citenamefont {Xiao}, \citenamefont {May},\ and\ \citenamefont
  {Bell}}]{Yang-JCED-2021-etarefrig}%
  \BibitemOpen
  \bibfield  {author} {\bibinfo {author} {\bibfnamefont {X.}~\bibnamefont
  {Yang}}, \bibinfo {author} {\bibfnamefont {X.}~\bibnamefont {Xiao}}, \bibinfo
  {author} {\bibfnamefont {E.~F.}\ \bibnamefont {May}},\ and\ \bibinfo {author}
  {\bibfnamefont {I.~H.}\ \bibnamefont {Bell}},\ }\bibfield  {title} {\bibinfo
  {title} {{Entropy Scaling of Viscosity{\textemdash}{III}: Application to
  Refrigerants and Their Mixtures}},\ }\href
  {https://doi.org/10.1021/acs.jced.0c01009} {\bibfield  {journal} {\bibinfo
  {journal} {J. Chem. Eng. Data}\ }\textbf {\bibinfo {volume} {66}},\ \bibinfo
  {pages} {1385} (\bibinfo {year} {2021})}\BibitemShut {NoStop}%
\bibitem [{\citenamefont {Young}\ \emph {et~al.}(2023)\citenamefont {Young},
  \citenamefont {Bell},\ and\ \citenamefont {Harvey}}]{Young-JCP-2023-salts}%
  \BibitemOpen
  \bibfield  {author} {\bibinfo {author} {\bibfnamefont {J.~M.}\ \bibnamefont
  {Young}}, \bibinfo {author} {\bibfnamefont {I.~H.}\ \bibnamefont {Bell}},\
  and\ \bibinfo {author} {\bibfnamefont {A.~H.}\ \bibnamefont {Harvey}},\
  }\bibfield  {title} {\bibinfo {title} {{Entropy Scaling of Viscosity for
  Molecular Models of Molten Salts}},\ }\href
  {https://doi.org/10.1063/5.0127250} {\bibfield  {journal} {\bibinfo
  {journal} {J. Chem. Phys.}\ }\textbf {\bibinfo {volume} {158}},\ \bibinfo
  {pages} {024502} (\bibinfo {year} {2023})}\BibitemShut {NoStop}%
\bibitem [{\citenamefont {Muzny}\ \emph {et~al.}(2013)\citenamefont {Muzny},
  \citenamefont {Huber},\ and\ \citenamefont {Kazakov}}]{Muzny-JCED-2013}%
  \BibitemOpen
  \bibfield  {author} {\bibinfo {author} {\bibfnamefont {C.~D.}\ \bibnamefont
  {Muzny}}, \bibinfo {author} {\bibfnamefont {M.~L.}\ \bibnamefont {Huber}},\
  and\ \bibinfo {author} {\bibfnamefont {A.~F.}\ \bibnamefont {Kazakov}},\
  }\bibfield  {title} {\bibinfo {title} {{Correlation for the Viscosity of
  Normal Hydrogen Obtained from Symbolic Regression}},\ }\bibfield  {journal}
  {\bibinfo  {journal} {J. Chem. Eng. Data}\ }\href
  {https://doi.org/10.1021/je301273j} {10.1021/je301273j} (\bibinfo {year}
  {2013})\BibitemShut {NoStop}%
\bibitem [{\citenamefont {Muzny}\ \emph {et~al.}(2022)\citenamefont {Muzny},
  \citenamefont {Huber},\ and\ \citenamefont {Kazakov}}]{Muzny-JCED-2022}%
  \BibitemOpen
  \bibfield  {author} {\bibinfo {author} {\bibfnamefont {C.~D.}\ \bibnamefont
  {Muzny}}, \bibinfo {author} {\bibfnamefont {M.~L.}\ \bibnamefont {Huber}},\
  and\ \bibinfo {author} {\bibfnamefont {A.~F.}\ \bibnamefont {Kazakov}},\
  }\bibfield  {title} {\bibinfo {title} {Erratum: Correlation for the viscosity
  of normal hydrogen obtained from symbolic regression},\ }\href
  {https://doi.org/10.1021/acs.jced.2c00523} {\bibfield  {journal} {\bibinfo
  {journal} {J. Chem. Eng. Data}\ }\textbf {\bibinfo {volume} {67}},\ \bibinfo
  {pages} {2855} (\bibinfo {year} {2022})}\BibitemShut {NoStop}%
\bibitem [{\citenamefont {Huber}(2018)}]{Huber-NISTIR-2018}%
  \BibitemOpen
  \bibfield  {author} {\bibinfo {author} {\bibfnamefont {M.~L.}\ \bibnamefont
  {Huber}},\ }\href {https://doi.org/10.6028/nist.ir.8209} {\emph {\bibinfo
  {title} {{Models for viscosity, thermal conductivity, and surface tension of
  selected pure fluids as implemented in {REFPROP} v10.0 (NISTIR 8209)}}}},\
  \bibinfo {type} {Tech. Rep.}\ (\bibinfo  {institution} {National Institute of
  Standards and Technology},\ \bibinfo {year} {2018})\BibitemShut {NoStop}%
\bibitem [{\citenamefont {Lemmon}\ \emph {et~al.}(2018)\citenamefont {Lemmon},
  \citenamefont {Bell}, \citenamefont {Huber},\ and\ \citenamefont
  {McLinden}}]{LEMMON-RP10}%
  \BibitemOpen
  \bibfield  {author} {\bibinfo {author} {\bibfnamefont {E.~W.}\ \bibnamefont
  {Lemmon}}, \bibinfo {author} {\bibfnamefont {I.~H.}\ \bibnamefont {Bell}},
  \bibinfo {author} {\bibfnamefont {M.~L.}\ \bibnamefont {Huber}},\ and\
  \bibinfo {author} {\bibfnamefont {M.~O.}\ \bibnamefont {McLinden}},\ }\href
  {https://doi.org/https://doi.org/10.18434/T4/1502528} {\bibinfo {title}
  {{NIST Standard Reference Database 23: Reference Fluid Thermodynamic and
  Transport Properties-REFPROP, Version 10.0, National Institute of Standards
  and Technology}}},\ \bibinfo {howpublished}
  {http://www.nist.gov/srd/nist23.cfm} (\bibinfo {year} {2018})\BibitemShut
  {NoStop}%
\bibitem [{\citenamefont {Rogers}\ and\ \citenamefont
  {Brickwedde}(1966)}]{Rogers-PHYSICA-1966}%
  \BibitemOpen
  \bibfield  {author} {\bibinfo {author} {\bibfnamefont {J.~D.}\ \bibnamefont
  {Rogers}}\ and\ \bibinfo {author} {\bibfnamefont {F.}~\bibnamefont
  {Brickwedde}},\ }\bibfield  {title} {\bibinfo {title} {Comparison of
  saturated-liquid viscosities of low molecular substances according to the
  quantum principle of corresponding states},\ }\href
  {https://doi.org/10.1016/0031-8914(66)90138-8} {\bibfield  {journal}
  {\bibinfo  {journal} {Physica}\ }\textbf {\bibinfo {volume} {32}},\ \bibinfo
  {pages} {1001} (\bibinfo {year} {1966})}\BibitemShut {NoStop}%
\bibitem [{\citenamefont {Leachman}\ \emph {et~al.}(2009)\citenamefont
  {Leachman}, \citenamefont {Jacobsen}, \citenamefont {Penoncello},\ and\
  \citenamefont {Lemmon}}]{10.1063/1.3160306}%
  \BibitemOpen
  \bibfield  {author} {\bibinfo {author} {\bibfnamefont {J.~W.}\ \bibnamefont
  {Leachman}}, \bibinfo {author} {\bibfnamefont {R.~T.}\ \bibnamefont
  {Jacobsen}}, \bibinfo {author} {\bibfnamefont {S.~G.}\ \bibnamefont
  {Penoncello}},\ and\ \bibinfo {author} {\bibfnamefont {E.~W.}\ \bibnamefont
  {Lemmon}},\ }\bibfield  {title} {\bibinfo {title} {Fundamental equations of
  state for parahydrogen, normal hydrogen, and orthohydrogen},\ }\href
  {https://doi.org/10.1063/1.3160306} {\bibfield  {journal} {\bibinfo
  {journal} {J. Phys. Chem. Ref. Data}\ }\textbf {\bibinfo {volume} {38}},\
  \bibinfo {pages} {721} (\bibinfo {year} {2009})}\BibitemShut {NoStop}%
\bibitem [{\citenamefont {Richardson}\ \emph {et~al.}(2014)\citenamefont
  {Richardson}, \citenamefont {Leachman},\ and\ \citenamefont
  {Lemmon}}]{10.1063/1.4864752}%
  \BibitemOpen
  \bibfield  {author} {\bibinfo {author} {\bibfnamefont {I.~A.}\ \bibnamefont
  {Richardson}}, \bibinfo {author} {\bibfnamefont {J.~W.}\ \bibnamefont
  {Leachman}},\ and\ \bibinfo {author} {\bibfnamefont {E.~W.}\ \bibnamefont
  {Lemmon}},\ }\bibfield  {title} {\bibinfo {title} {Fundamental equation of
  state for deuterium},\ }\href {https://doi.org/10.1063/1.4864752} {\bibfield
  {journal} {\bibinfo  {journal} {J. Phys. Chem. Ref. Data}\ }\textbf {\bibinfo
  {volume} {43}},\ \bibinfo {pages} {013103} (\bibinfo {year}
  {2014})}\BibitemShut {NoStop}%
\bibitem [{\citenamefont {Tegeler}\ \emph {et~al.}(1999)\citenamefont
  {Tegeler}, \citenamefont {Span},\ and\ \citenamefont
  {Wagner}}]{10.1063/1.556037}%
  \BibitemOpen
  \bibfield  {author} {\bibinfo {author} {\bibfnamefont {C.}~\bibnamefont
  {Tegeler}}, \bibinfo {author} {\bibfnamefont {R.}~\bibnamefont {Span}},\ and\
  \bibinfo {author} {\bibfnamefont {W.}~\bibnamefont {Wagner}},\ }\bibfield
  {title} {\bibinfo {title} {A new equation of state for argon covering the
  fluid region for temperatures from the melting line to 700 k at pressures up
  to 1000 {MPa}},\ }\href {https://doi.org/10.1063/1.556037} {\bibfield
  {journal} {\bibinfo  {journal} {J. Phys. Chem. Ref. Data}\ }\textbf {\bibinfo
  {volume} {28}},\ \bibinfo {pages} {779} (\bibinfo {year} {1999})}\BibitemShut
  {NoStop}%
\bibitem [{\citenamefont {Lemmon}\ and\ \citenamefont
  {Span}(2006)}]{10.1021/je050186n}%
  \BibitemOpen
  \bibfield  {author} {\bibinfo {author} {\bibfnamefont {E.~W.}\ \bibnamefont
  {Lemmon}}\ and\ \bibinfo {author} {\bibfnamefont {R.}~\bibnamefont {Span}},\
  }\bibfield  {title} {\bibinfo {title} {Short fundamental equations of state
  for 20 industrial fluids},\ }\href {https://doi.org/10.1021/je050186n}
  {\bibfield  {journal} {\bibinfo  {journal} {J. Chem. Eng. Data}\ }\textbf
  {\bibinfo {volume} {51}},\ \bibinfo {pages} {785} (\bibinfo {year}
  {2006})}\BibitemShut {NoStop}%
\bibitem [{\citenamefont {Arp}\ \emph {et~al.}(1998)\citenamefont {Arp},
  \citenamefont {McCarty},\ and\ \citenamefont {Friend}}]{Arp-NIST-1998}%
  \BibitemOpen
  \bibfield  {author} {\bibinfo {author} {\bibfnamefont {V.}~\bibnamefont
  {Arp}}, \bibinfo {author} {\bibfnamefont {R.}~\bibnamefont {McCarty}},\ and\
  \bibinfo {author} {\bibfnamefont {D.}~\bibnamefont {Friend}},\ }\href@noop {}
  {\emph {\bibinfo {title} {{Thermophysical Properties of Helium-4 from 0.8 to
  1500 K with Pressures to 2000 MPa - NIST Technical Note 1334 (revised)}}}},\
  \bibinfo {type} {Tech. Rep.}\ (\bibinfo  {institution} {NIST},\ \bibinfo
  {year} {1998})\BibitemShut {NoStop}%
\bibitem [{\citenamefont {{da C. Andrade}}(1931)}]{Andrade-NATURE-1931}%
  \BibitemOpen
  \bibfield  {author} {\bibinfo {author} {\bibfnamefont {E.~N.}\ \bibnamefont
  {{da C. Andrade}}},\ }\bibfield  {title} {\bibinfo {title} {{Viscosity of
  Liquids}},\ }\href {https://doi.org/10.1038/128835a0} {\bibfield  {journal}
  {\bibinfo  {journal} {Nature}\ }\textbf {\bibinfo {volume} {128}},\ \bibinfo
  {pages} {835} (\bibinfo {year} {1931})}\BibitemShut {NoStop}%
\bibitem [{\citenamefont {Chapman}\ and\ \citenamefont
  {Cowling}(1970)}]{Chapman-BOOK-1970}%
  \BibitemOpen
  \bibfield  {author} {\bibinfo {author} {\bibfnamefont {S.}~\bibnamefont
  {Chapman}}\ and\ \bibinfo {author} {\bibfnamefont {T.~G.}\ \bibnamefont
  {Cowling}},\ }\href@noop {} {\emph {\bibinfo {title} {{The Mathematical
  Theory of Non-uniform Gases: An Account of the Kinetic Theory of Viscosity,
  Thermal Conduction and Diffusion in Gases}}}}\ (\bibinfo  {publisher}
  {Cambridge University Press},\ \bibinfo {year} {1970})\BibitemShut {NoStop}%
\bibitem [{\citenamefont {Aasen}\ \emph {et~al.}(2019)\citenamefont {Aasen},
  \citenamefont {Hammer}, \citenamefont {Ervik}, \citenamefont {Müller},\ and\
  \citenamefont {Wilhelmsen}}]{Aasen-JCP-2019}%
  \BibitemOpen
  \bibfield  {author} {\bibinfo {author} {\bibfnamefont {A.}~\bibnamefont
  {Aasen}}, \bibinfo {author} {\bibfnamefont {M.}~\bibnamefont {Hammer}},
  \bibinfo {author} {\bibfnamefont {{\AA}.}~\bibnamefont {Ervik}}, \bibinfo
  {author} {\bibfnamefont {E.~A.}\ \bibnamefont {Müller}},\ and\ \bibinfo
  {author} {\bibfnamefont {{\O}.}~\bibnamefont {Wilhelmsen}},\ }\bibfield
  {title} {\bibinfo {title} {{Equation of state and force fields for
  Feynman-Hibbs-corrected Mie fluids. I. Application to pure helium, neon,
  hydrogen, and deuterium}},\ }\href {https://doi.org/10.1063/1.5111364}
  {\bibfield  {journal} {\bibinfo  {journal} {J. Chem. Phys.}\ }\textbf
  {\bibinfo {volume} {151}},\ \bibinfo {pages} {064508} (\bibinfo {year}
  {2019})}\BibitemShut {NoStop}%
\bibitem [{\citenamefont {Aasen}\ \emph
  {et~al.}(2020{\natexlab{a}})\citenamefont {Aasen}, \citenamefont {Hammer},
  \citenamefont {M{\"u}ller},\ and\ \citenamefont
  {Wilhelmsen}}]{Aasen-JCP-2020}%
  \BibitemOpen
  \bibfield  {author} {\bibinfo {author} {\bibfnamefont {A.}~\bibnamefont
  {Aasen}}, \bibinfo {author} {\bibfnamefont {M.}~\bibnamefont {Hammer}},
  \bibinfo {author} {\bibfnamefont {E.~A.}\ \bibnamefont {M{\"u}ller}},\ and\
  \bibinfo {author} {\bibfnamefont {{\O}.}~\bibnamefont {Wilhelmsen}},\
  }\bibfield  {title} {\bibinfo {title} {{Equation of state and force fields
  for Feynman-Hibbs-corrected Mie fluids. {II}. Application to mixtures of
  helium, neon, hydrogen, and deuterium}},\ }\href
  {https://doi.org/10.1063/1.5136079} {\bibfield  {journal} {\bibinfo
  {journal} {J. Chem. Phys.}\ }\textbf {\bibinfo {volume} {152}},\ \bibinfo
  {pages} {074507} (\bibinfo {year} {2020}{\natexlab{a}})}\BibitemShut
  {NoStop}%
\bibitem [{\citenamefont {Aasen}\ \emph
  {et~al.}(2020{\natexlab{b}})\citenamefont {Aasen}, \citenamefont {Hammer},
  \citenamefont {Lasala}, \citenamefont {Jaubert},\ and\ \citenamefont
  {Wilhelmsen}}]{Aasen-FPE-2020}%
  \BibitemOpen
  \bibfield  {author} {\bibinfo {author} {\bibfnamefont {A.}~\bibnamefont
  {Aasen}}, \bibinfo {author} {\bibfnamefont {M.}~\bibnamefont {Hammer}},
  \bibinfo {author} {\bibfnamefont {S.}~\bibnamefont {Lasala}}, \bibinfo
  {author} {\bibfnamefont {J.-N.}\ \bibnamefont {Jaubert}},\ and\ \bibinfo
  {author} {\bibfnamefont {{\O}.}~\bibnamefont {Wilhelmsen}},\ }\bibfield
  {title} {\bibinfo {title} {Accurate quantum-corrected cubic equations of
  state for helium, neon, hydrogen, deuterium and their mixtures},\ }\href
  {https://doi.org/10.1016/j.fluid.2020.112790} {\bibfield  {journal} {\bibinfo
   {journal} {Fluid Phase Equilibria}\ }\textbf {\bibinfo {volume} {524}},\
  \bibinfo {pages} {112790} (\bibinfo {year} {2020}{\natexlab{b}})}\BibitemShut
  {NoStop}%
\bibitem [{\citenamefont {Walker}\ \emph {et~al.}(2022)\citenamefont {Walker},
  \citenamefont {Zhao}, \citenamefont {Haslam},\ and\ \citenamefont
  {Jackson}}]{Walker-JCP-2022}%
  \BibitemOpen
  \bibfield  {author} {\bibinfo {author} {\bibfnamefont {P.~J.}\ \bibnamefont
  {Walker}}, \bibinfo {author} {\bibfnamefont {T.}~\bibnamefont {Zhao}},
  \bibinfo {author} {\bibfnamefont {A.~J.}\ \bibnamefont {Haslam}},\ and\
  \bibinfo {author} {\bibfnamefont {G.}~\bibnamefont {Jackson}},\ }\bibfield
  {title} {\bibinfo {title} {{Ab initio development of generalized
  Lennard-Jones (Mie) force fields for predictions of thermodynamic properties
  in advanced molecular-based {SAFT} equations of state}},\ }\href
  {https://doi.org/10.1063/5.0087125} {\bibfield  {journal} {\bibinfo
  {journal} {J. Chem. Phys.}\ }\textbf {\bibinfo {volume} {156}},\ \bibinfo
  {pages} {154106} (\bibinfo {year} {2022})}\BibitemShut {NoStop}%
\bibitem [{\citenamefont {Yoon}\ and\ \citenamefont
  {Scheraga}(1988)}]{Yoon-JCP-1988}%
  \BibitemOpen
  \bibfield  {author} {\bibinfo {author} {\bibfnamefont {B.}~\bibnamefont
  {Yoon}}\ and\ \bibinfo {author} {\bibfnamefont {H.~A.}\ \bibnamefont
  {Scheraga}},\ }\bibfield  {title} {\bibinfo {title} {{Monte Carlo simulation
  of the hard‐sphere fluid with quantum correction and estimate of its free
  energy}},\ }\href {https://doi.org/10.1063/1.453841} {\bibfield  {journal}
  {\bibinfo  {journal} {J. Chem. Phys.}\ }\textbf {\bibinfo {volume} {88}},\
  \bibinfo {pages} {3923} (\bibinfo {year} {1988})}\BibitemShut {NoStop}%
\bibitem [{\citenamefont {Deiters}(1983)}]{Deiters-FPE-1983}%
  \BibitemOpen
  \bibfield  {author} {\bibinfo {author} {\bibfnamefont {U.~K.}\ \bibnamefont
  {Deiters}},\ }\bibfield  {title} {\bibinfo {title} {Calculation and
  prediction of fluid phase equilibria from an equation of state},\ }\href
  {https://doi.org/https://doi.org/10.1016/0378-3812(83)80032-6} {\bibfield
  {journal} {\bibinfo  {journal} {Fluid Phase Equilib.}\ }\textbf {\bibinfo
  {volume} {10}},\ \bibinfo {pages} {173} (\bibinfo {year} {1983})}\BibitemShut
  {NoStop}%
\bibitem [{\citenamefont {Deiters}\ and\ \citenamefont
  {Sadus}(2020)}]{Deiters-JPCB-2020}%
  \BibitemOpen
  \bibfield  {author} {\bibinfo {author} {\bibfnamefont {U.~K.}\ \bibnamefont
  {Deiters}}\ and\ \bibinfo {author} {\bibfnamefont {R.~J.}\ \bibnamefont
  {Sadus}},\ }\bibfield  {title} {\bibinfo {title} {Ab initio interatomic
  potentials and the classical molecular simulation prediction of the
  thermophysical properties of helium},\ }\href
  {https://doi.org/10.1021/acs.jpcb.9b11108} {\bibfield  {journal} {\bibinfo
  {journal} {J. Phys. Chem. B}\ }\textbf {\bibinfo {volume} {124}},\ \bibinfo
  {pages} {2268} (\bibinfo {year} {2020})}\BibitemShut {NoStop}%
\bibitem [{\citenamefont {Deiters}\ and\ \citenamefont
  {Sadus}(2021)}]{Deiters-JPCB-2021}%
  \BibitemOpen
  \bibfield  {author} {\bibinfo {author} {\bibfnamefont {U.~K.}\ \bibnamefont
  {Deiters}}\ and\ \bibinfo {author} {\bibfnamefont {R.~J.}\ \bibnamefont
  {Sadus}},\ }\bibfield  {title} {\bibinfo {title} {Interatomic interactions
  responsible for the solid-liquid and vapor-liquid phase equilibria of neon},\
  }\href {https://doi.org/10.1021/acs.jpcb.1c04272} {\bibfield  {journal}
  {\bibinfo  {journal} {J. Phys. Chem. B}\ }\textbf {\bibinfo {volume} {125}},\
  \bibinfo {pages} {8522} (\bibinfo {year} {2021})}\BibitemShut {NoStop}%
\bibitem [{\citenamefont {Sadus}\ and\ \citenamefont
  {Deiters}(tted)}]{Sadus-JCP-2023}%
  \BibitemOpen
  \bibfield  {author} {\bibinfo {author} {\bibfnamefont {R.}~\bibnamefont
  {Sadus}}\ and\ \bibinfo {author} {\bibfnamefont {U.}~\bibnamefont
  {Deiters}},\ }\bibfield  {title} {\bibinfo {title} {{An intermolecular
  potential for normal hydrogen: Classical molecular simulation of vapor-liquid
  equilibria, critical and triple point properties}},\ }\href@noop {}
  {\bibfield  {journal} {\bibinfo  {journal} {J. Chem. Phys.}\ } (\bibinfo
  {year} {2023 (submitted)})}\BibitemShut {NoStop}%
\bibitem [{\citenamefont {Herreman}\ and\ \citenamefont
  {Grevendonk}(1974)}]{Herreman-CRYO-1974}%
  \BibitemOpen
  \bibfield  {author} {\bibinfo {author} {\bibfnamefont {W.}~\bibnamefont
  {Herreman}}\ and\ \bibinfo {author} {\bibfnamefont {W.}~\bibnamefont
  {Grevendonk}},\ }\bibfield  {title} {\bibinfo {title} {An experimental study
  on the shear viscosity of liquid neon},\ }\href
  {https://doi.org/10.1016/0011-2275(74)90081-2} {\bibfield  {journal}
  {\bibinfo  {journal} {Cryogenics}\ }\textbf {\bibinfo {volume} {14}},\
  \bibinfo {pages} {395} (\bibinfo {year} {1974})}\BibitemShut {NoStop}%
\bibitem [{\citenamefont {Rutkai}\ \emph {et~al.}(2017)\citenamefont {Rutkai},
  \citenamefont {Thol}, \citenamefont {Span},\ and\ \citenamefont
  {Vrabec}}]{Rutkai-MP-2016}%
  \BibitemOpen
  \bibfield  {author} {\bibinfo {author} {\bibfnamefont {G.}~\bibnamefont
  {Rutkai}}, \bibinfo {author} {\bibfnamefont {M.}~\bibnamefont {Thol}},
  \bibinfo {author} {\bibfnamefont {R.}~\bibnamefont {Span}},\ and\ \bibinfo
  {author} {\bibfnamefont {J.}~\bibnamefont {Vrabec}},\ }\bibfield  {title}
  {\bibinfo {title} {{How well does the Lennard-Jones potential represent the
  thermodynamic properties of noble gases?}},\ }\href
  {https://doi.org/10.1080/00268976.2016.1246760} {\bibfield  {journal}
  {\bibinfo  {journal} {Mol. Phys.}\ }\textbf {\bibinfo {volume} {115}},\
  \bibinfo {pages} {1104} (\bibinfo {year} {2017})}\BibitemShut {NoStop}%
\bibitem [{\citenamefont {Nagashima}\ \emph {et~al.}(2017)\citenamefont
  {Nagashima}, \citenamefont {Tsuda}, \citenamefont {Tsuboi}, \citenamefont
  {Hayashi},\ and\ \citenamefont {Tokumasu}}]{Nagashima-JCP-2017}%
  \BibitemOpen
  \bibfield  {author} {\bibinfo {author} {\bibfnamefont {H.}~\bibnamefont
  {Nagashima}}, \bibinfo {author} {\bibfnamefont {S.}~\bibnamefont {Tsuda}},
  \bibinfo {author} {\bibfnamefont {N.}~\bibnamefont {Tsuboi}}, \bibinfo
  {author} {\bibfnamefont {A.~K.}\ \bibnamefont {Hayashi}},\ and\ \bibinfo
  {author} {\bibfnamefont {T.}~\bibnamefont {Tokumasu}},\ }\bibfield  {title}
  {\bibinfo {title} {A molecular dynamics study of nuclear quantum effect on
  diffusivity of hydrogen molecule},\ }\href
  {https://doi.org/10.1063/1.4991732} {\bibfield  {journal} {\bibinfo
  {journal} {J. Chem. Phys.}\ }\textbf {\bibinfo {volume} {147}},\ \bibinfo
  {pages} {024501} (\bibinfo {year} {2017})}\BibitemShut {NoStop}%
\end{thebibliography}
\end{document}